\documentstyle[aps,epsf]{revtex}
\begin{document}
\title{Evidence for Negative Stiffness of QCD Strings}
\author{H.~Kleinert}
\address{Institut f\"ur Theoretische Physik\\
Freie Universit\"at Berlin, Arnimallee 14, 1000 Berlin 33, Germany}
\author{A.M. Chervyakov}
\address{Joint Institut for Nuclear Research, Dubna, SU 141980, Russia}
\date{\today}
\maketitle
\begin{abstract}
QCD strings are color-electric flux tubes between quarks
with a finite thickness determined by the dimensionally
transmuted coupling constant, and thus with
a finite
curvature stiffness. Contrary to an earlier
rigid-string model
by Polyakov and Kleinert,
and motivated by the properties of a magnetic flux tubes
in type-II superconductors, we
put forward the hypothesis that
QCD strings have a {\em negative\/} stiffness.
We set up a new string
model with this property and show that it is free of
the three principal problems of
rigid strings --- particle states with negative norm,
nonexistence of a lowest-energy state, and wrong
high-temperature behavior of string tension ---
thus making it a better candidate for a string
description of quark forces than previous models.
\\
PACS number(s): 11.17.+y, 12.38.Aw, 12.40.Aa \\ ~\\
\end{abstract}

Color-electric flux tubes between quarks have a thickness
of the order of the dimensionally transmuted
coupling constant. They should therefore display a finite
resistance to a change in extrinsic curvature.
To account for
this property, a higher-derivative
stiffness term was added
to the Nambu-Goto (NG) action of an infinitely
thin string \cite{p:rstring,k:rstring}. The resulting
Polyakov-Kleinert (PK)
string
seemed to render a more relevant description
of gluon
forces in QCD than the former NG string.
Particularly attractive was the fact that a PK model
containing {\em only\/} a stiffness term is asymptotically
free at short distances \cite{p:rstring,k:rstring}
and can generate the string tension
spontaneously \cite{k:sptension}.
For other
properties see \cite{kg:atrprop}.

Unfortunately,  however, the PK model has also serious
consistency problems.
First of all, there exists
an unphysical ghost pole in the propagator
\cite{k:rstring}. The pole is generated
by the second derivatives with respect to the string
coordinates in the action.
Second, the PK model shares with all higher-derivative
theories
the lack of a lowest-energy state
\cite{h:hder,s:nonlocal,bz:lackenergy,chn:hder}.
Third, the squared
string tension of the PK model has
an unphysical imaginary part at high
temperatures or low $ \beta=1/T$, even though this model
unlike the
NG string
displays the correct power
behavior  $ \beta^{-2}$ for $ \beta\rightarrow 0$
consistent with asymptotic freedom
(except, however, for an imaginary factor $i$) \cite{py:htemp}.

It is the purpose of this letter
to propose a new string model in which the problems of the PK string
are
absent while
the
attractive features are preserved.
This new model has a {\em negative\/} extrinsic curvature stiffness.
It was inspired by properties of magnetic flux tubes
in type-II superconductors and
analogous properties of Nielsen-Olesen vortices in
relativistic gauge models.
In the London limit,
these have a nontrivial energy spectrum, from which
several calculations have deduced
a negative stiffness
\cite{k:negstiff,NOvortex,kk:biomembr}.
By analogy, we
hypothesize that the stiffness QCD strings
is also {\it negative}.  The action we propose for
a description of such a string is
\begin{equation}
{\cal A}\,=\,\frac{c-1}{2}M^{2}\,
\int\limits_{}^{}\,d^{2}\xi\sqrt{g}\,g^{ij}\,D_{i}x^{\lambda}\,
\frac{1}{c\,-\,e^{D^{2}/\mu^2}}\,D_{j}x_{\lambda}\,{,}
\label{act}\end{equation}
where $x^{\lambda}(\xi)$ with $\lambda\,=\,0,1,\ldots,d-1$
are the string coordinates in a $d$-dimensional
euclidean spacetime parametrized by $\xi^{i},\,i\,=\,0,1$,
and $g_{ij}= \partial _ix^ \lambda\partial _jx_ \lambda$ is the induced
metric on the
world surface of the
string  with $g^{ij}$ being its inverse and $g\,=\,\det(g_{ij})$.
Covariant differentiation with respect to $\xi^{i}$ is denoted by $D_{i}$,
and $D^{2}\,=\,D_{i}D^{i}$
is Laplace-Beltrami operator. The physics of this model
is governed by two mass scales $M$ and
$\mu$, and a
dimensionless constant  $c$.

Due to nonzero radius of thickness the interaction between surface elements
of the string has a nonlocal character, just like in a magnetic flux tube.
In momentum space, the quadratic part of the action (\ref{act})
gives rise to the free propagator
\begin{equation}
G(k^2)\,=\,\frac{1}{c-1}\frac{c\,-\,e^{-k^{2}/\mu^2}}{k^2}.
\label{green}\end{equation}
For small $k$, this has an expansion
\begin{equation}
G(k^2)\,=\,\frac{1+k^2/ \Lambda^2}{k^2} +\dots,
\label{greenapp}\end{equation}
with the mass parameter $ \Lambda^2\equiv (c-1)\mu^2$.
The stiffness $ \alpha$ is the inverse coefficient of $k^4/2$
in the action  and has the negative
value
$\alpha\,=\,-\Lambda^2/M^2$.

The propagator contains only a single pole at $k^2\,=\,0$.
This is to be contrasted with
the PK model
where
\begin{equation}
G(k^2)\,=\,\frac{1}{k^2(1+k^2/ \Lambda^2)}
\label{greenrs}\end{equation}
corresponds to the positive stiffness
$\alpha\,=\,\Lambda^2/M^2$
and
contains an
unphysical pole with negative residue at $k^2=- \Lambda^2$.

\begin{figure}[tb]
\unitlength=1mm
\def\fsz{\footnotesize}
\def\ssz{\scriptsize}
\def\tsz{\tiny}
\def\pu#1#2{\put(#1,#2){\emmoveto}}
\def\pd#1#2{\put(#1,#2){\emlineto}}
\begin{picture}(78.48,53.25)
\def\IncludeEpsImg#1#2#3#4{\renewcommand{\epsfsize}[2]{#3##1}{\epsfbox{#4}}}
\centerline{
\put(-79.24,-5){\IncludeEpsImg{78.48mm}{53.25mm}{1.0000}{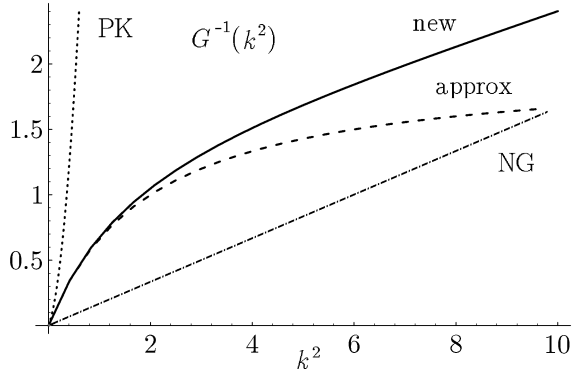}}
}
\end{picture}

{}~\\
\caption[]{\footnotesize
The inverse propagators
of the model (\ref{act}) (solid curve) and of the approximation
(\ref{act1})
(dashed lines) for $\mu\,=\, 10$
and $c=1.2$. The dotted curve shows the
behavior in a PK string
with the opposite value of the curvature stiffness, the dash-dotted
curve corresponds to a NG string with
reduced tension $(c-1)M^2/c$. We see that the large-$k$
fluctuations are much
more violent than
in the PK string, but less violent than
in a NG string, which is eventually approached.
}
\label{fig1}
\end{figure}

The inverse propagator (\ref{green})
is plotted in Fig.~\ref{fig1} for sample parameters
$\mu$ and $c$. The
fluctuations with large momenta are more violent than in the PK
string with the opposite
curvature stiffness, but less violent than in
a NG string with reduced tension $(c-1)M^2/c$,
which is eventually approached.

The above string is inspired by the properties
of
a vortex line in a type-II superconductor \cite{k:gfields},
which
has a diameter of the order of
the penetration depth
governing
the parameter
$1/\Lambda$ of the model.
The inverse mass $1/\mu$ corresponds to the coherence length
which goes to zero in the London limit.

In QCD, the parameters of the model have the
following physical origin.
As long as there are no
quarks, QCD gives rise to
a string with only one length scale
which enters into both the string tension
and the thickness of the flux tube.
Thus it determines the two parameters
$M$ and $ \Lambda$ with some ratio
between them, which
is a consequence
of the detailed dynamics of the gluon system.
As quarks are included,
this ratio is modified
and an additional length scale arises
at which the
vacuum polarization of the gluon propagator
becomes important
and changes the stiffness
properties of the flux tube.
This length scale is proportional
to some average over
 the inverse quark masses,
and
accounted for by
the model parameter
$1/\mu$.

For practical calculations,
it is convenient
to approximate (\ref{green}) by
(\ref{greenapp}),
and cut all
momentum integrals off at $k^2=\mu^2$.
To have a physical penetration depth, we must take $\Lambda$
to be smaller than
the cutoff $\mu$, or
$1<c<2$.

The action associated with the approximate Green
function (\ref{greenapp}) is
\begin{equation}
{\cal A}_1\,=\,\frac{1}{2}M^{2}\,
\int\limits_{}^{}\,d^{2}\xi\sqrt{g}\,g^{ij}\,D_{i}x^{\lambda}\,
\frac{1}{1-D^2/ \Lambda^2}\,D_{j}x_{\lambda}\,{.}
\label{act1}\end{equation}

We shall now  derive the high-temperature limit of the total
string tension, which can be done exactly in
the limit $d\rightarrow \infty$.
As usual in this limit,
we make the
metric field $g_{ij}$ an independent fluctuating field,
and force it to be equal
to the induced metric $\partial_{i}x^{\lambda}\partial_{j}x_{\lambda}$
by means of
Lagrange multipliers $\lambda^{ij}$, adding to (\ref{act1})
a term
\begin{equation} {\cal A}_{2}\,
=\,\frac{1}{2}M^{2}\int\limits^{}_{}d^{2}\xi\sqrt{g}\,
\lambda^{ij}\,\left(\partial_{i}x^{\lambda}
\partial_{j}x_{\lambda}\,-\,g_{ij}\right)\,{.}
\label{act2}\end{equation}
It is straight-forward to calculate the partition function
of the string at a finite temperature $T$ anchored to two static quarks
separated by a
large distance $R$. The world sheet
of the string
is periodic in the imaginary-time direction with a
period $\beta\,=\,T^{-1}$. Choosing
coordinates with
$\xi^{1}$ running from $0$ to $R$,
and $\xi^{0}$ from $0$ to $\beta$, we parametrize
the world sheet as
$$
x^{\lambda}(\xi)\,=\,\left(\xi^{0},\,\xi^{1},\,x^{a}(\xi)\right),
$$
with $x^a(\xi),~
a\,=\,2,3,\ldots\,d-1,$
describing the transverse fluctuations
which occur quadratically in the extended action
${\cal A}_{1}+{\cal A}_{2}$.
These can be integrated out producing a trace log with a factor $d-2$
which in the limit
$d\rightarrow \infty$ suppresses the fluctuations
of $g_{ij}$ and $ \lambda^{ij}$,
so that the saddle point approximation provides us with
an exact partition function.
At the saddle point, we may
assume diagonal forms for
the metric and the Lagrange multipliers:
\begin{equation}
g_{ij}\,=\,\rho_{i}\delta_{ij},\quad \lambda^{ij}\,=\,\lambda_{i}g^{ij},
\label{matrices}\end{equation}
with the constants $\rho_{i}$ and $\lambda_{i}$.
The resulting effective action
${\cal A}_{\rm eff}$
=${\cal A}_{\rm eff}^{\rm mf}$+
${\cal A}_{\rm eff}^{\rm loop}$
consists of a mean-field term
\begin{equation}
{\cal A}_{\rm eff}^{\rm mf}\,
=\,\frac{R\beta M^2}{2}\,
\sqrt{\rho_{0}\rho_{1}}\,\left[\left(\frac{\lambda_0\,+\,1}{\rho_0}\,
+\,\frac{\lambda_1\,+\,1}{\rho_1}\right)\,-\,
\left(\lambda_0\,+\lambda_1\right)\right],
\label{mfact}\end{equation}
which also appears in NG and PK string models,
and a one-loop contribution
\begin{eqnarray}
&&\,
{\cal A}_{\rm eff}^{\rm loop}\,=\,\frac{d-2}{2}\,R\,\sqrt{\rho_{1}}\,\sum
\limits_{-\infty}^{\infty}\int\limits_{}^{}
\frac{dq_{1}}{2\pi}\,\ln\Biggl\{\left[\frac{4\pi^{2}n^2}
{\rho_{0}\beta^{2}}\,+\,q^{2}_{1}\right]\left[
1\,+\,\frac{1}{ \Lambda^2}\left(\frac{4\pi^{2}n^2}
{\rho_{0}\beta^2}\,+\,q^{2}_{1}\right)\right]^{-1}
\nonumber\\
&&~~~~~~~~~~~~~~~~~~~~~~~~~~~~~~~~~~~~~~~~~~~~~~~+\,\lambda_0
\,\frac{4\pi^{2}n^2}
{\rho_{0}\beta^2}\,+\,\lambda_1
\,q^{2}_{1}\Biggr\}\,{.}
\label{loopact}\end{eqnarray}
Here
the temporal
components $q_0$ are summed over all thermal Matsubara frequencies
$q_0\,= (q_0)_n\equiv 2\pi n/\beta\sqrt{\rho_0},\quad
n\,=\,0,\pm 1,\ldots\,~$.

The effective string tension is defined by
$M^2_{\rm eff}\,=\,F/\beta$, where $ F=
{\cal A}_{\rm eff}/R$ is the free energy per unit length
evaluated at the extremal values of $\rho_i$ and $\lambda_i$.
At low temperatures,
$\beta\,\gg\,1/ \Lambda$ and the one-loop term (\ref{loopact}) is
very small compared to the mean-field term (\ref{mfact}),
so that the
extremum of ${\cal A}_{\rm eff}$ lies at
$\rho_0\,=\,\rho_1\,=\,1$,~$\lambda_0\,=\,\lambda_1\,=\,0$, where
$F\,=\,\beta M^2$.  At high temperatures, on the other hand,
$\beta\,\ll\,1/ \Lambda$
and the one-loop term (\ref{loopact}) becomes
essential. In this limit, it can easily be evaluated analytically.

First, however, let us make
an instructive
observation:
If we assume for a moment that the cutoff $\mu^2$ be artificially small,
so that
$q_0^2+q_1^2\ll  \Lambda^2$, we can neglect
the denominator of the first term in Eq.~(\ref{loopact}) and see
that
the reduced one-loop action (\ref{loopact})
takes the
NG form.
By the same calculation as in
Ref.\ \onlinecite{pa:NGstring}
we then find at high temperatures
the squared free energy
per unit length
\begin{equation}
F^2\approx F^2_{{\rm NG}}(\beta)\,=\,\beta^2M^4
\left[1\,-\,\frac{(d-2)\pi}{3M^2\beta^2}\right]\,\approx\,
-\frac{(d-2)\pi M^2}{3}.
\label{NGenergy}\end{equation}
This is in bad qualitative disagreement
with the behavior derived from
QCD in the limit of
large-$N$ by Polchinski \cite{p:QCDstring}:
\begin{equation}
F^2_{{\rm QCD}}(\beta)\,\approx\,-\frac{2g^{2}(\beta)N}{\pi^{2}\beta^2}.
\label{QCDenergy}\end{equation}

In contrast to this, the present
model has the correct high-temperature behavior. Since the
cutoff $\mu^2$ is, of course, much
larger than $\Lambda^2$,
the negative curvature stiffness
$\alpha\,=\,-\Lambda^2/M^2$ can take effect.
In the high-temperature limit $\beta\,\ll\,1/\Lambda$,
the first term in the logarithm (\ref{loopact})
is small compared to the second-derivative term,
and one-loop
action (\ref{loopact}) reduces to
\begin{equation}
{\cal A}_{\rm eff}^{\rm loop}\,\approx\,\frac{d-2}{2}\,R\,
\sqrt{\frac{\rho_1}{\lambda_1}}
\left(\Lambda\,-\,\frac{\pi}{3\beta}
\,\sqrt{\frac{\lambda_0}{\rho_0}}\right)\,{.}
\label{loopactapp}\end{equation}
The first term comes only from zero Matsubara frequency,
in the large $\mu^2$-limit,
the second contains the
full spectral sum in (\ref{loopact}), but approximated by the
leading second-derivative term.
Adding to (\ref{loopactapp})
the mean-field term (\ref{mfact}),
the resulting
high-temperature
effective
action ${\cal A}_{\rm eff}$ has a saddle point at
\begin{equation}
\rho_{0}^{-1}\,=\,2\,-\,\lambda_1\,\frac{\lambda_0\,-\,\lambda_1}
{\lambda_0\,+\,\lambda_1}\,{,}\quad
\rho_{1}^{-1}\,=\,2\,+\,\lambda_0\,\frac{\lambda_0\,-\,\lambda_1}
{\lambda_0\,+\,\lambda_1}\,{,}
\label{extr1}\end{equation}
where $\lambda_0$ and $\lambda_1$ are determined from
\begin{equation}
\frac{2\beta M^2}{(d-2)\Lambda}\,\sqrt{\rho_0\lambda_1}\,
\left(\lambda_0\,+\,\lambda_1\right)\,=\,1\,{,}\quad
\sqrt{\lambda_0\lambda_1}\,\left(\rho_0^{-1}\,-\,1\right)\,=\,
\frac{(d-2)\pi}{6\beta^{2}M^2}\,\rho_0^{-1}\,{.}
\label{extr2}\end{equation}
In terms of $\lambda_1$, the associated squared free
energy per unit length reads simply
\begin{equation}
F^2\,=\,M^{4}\beta^2\frac{\rho_0}{\rho_1}
\left(\lambda_1\,+\,1\right)^2\,{.}
\label{energy}\end{equation}
The deconfinement temperature $T_{\rm d}$ is determined by
the vanishing of $1/\rho_1$, which yields
\begin{equation}
\frac{T_{\rm d}}{M_{\rm tot}(0)}\,=\,
\frac{2}{(d-2)}\left[b(|\alpha|)|\alpha|\left(1\,+\,
\sqrt{\frac{(d-2)|\alpha|}{8\pi}}\right)\right]^{-1/2}\,{.}
\label{dtemp}\end{equation}
Here $M_{\rm tot}(0)\,=\,M\,\sqrt{\nu\,+\,1}$ is the total
string tension of at infinite $beta$, and
the parameter $\nu$ is related to the stiffness $\alpha$
as follows
$$
\nu\,=\,\sqrt{\frac{(d-2)|\alpha|}{8\pi}}\,{.}
$$
The parameter $b(|\alpha|)$ is an unique real positive root
of the polynomial equation of 9th order which
exists for a reasonable range of the mass parameter
$\Lambda\,\geq\,8.25M$. Numerical studies
show that the deconfinement temperature (\ref{dtemp})
depends only weakly on the stiffness parameter $\alpha$
and lies in our model always around
$T_{\rm d}/M_{\rm tot}(0)\,\approx\,0.932$.
This is somewhat larger
than the estimates
derived from rigid strings \cite{kg:atrprop}.

Equations
(\ref{extr1}) and (\ref{extr2})
determine
$\lambda_1$
via a polynomial of 14th order. In the above
range of the mass parameter, $\Lambda\,\geq\,8.25M$,
there
are
real roots for all $ \beta$.
At high-temperatures,
these have the asymptotic behavior
\begin{equation}
\lambda_{1}\,={\rm const.}\times\beta^0{,}~~~
|{\rm const.}|\,>\,2\,{;}\quad
{}~~~~~\lambda_{0}\,=\,\frac{(d-2)^2\pi^2}{36M^4\beta^4}\,
\frac{(\lambda_1\,-\,2)^2}{\lambda_{1}(\lambda_1\,-\,1)^2}\,{.}
\label{roots}\end{equation}
Inserted into
(\ref{extr1}), they yield
the diagonal elements of the metric.
The resulting high-temperature behavior
of (\ref{energy})
is, therefore,
\begin{eqnarray}
F^2&\approx&
-\,\frac{\pi^{2}(d-2)^2}{36\beta^2}\,
\frac{(\lambda_1\,-\,2)(\lambda_1\,+\,1)^2}
{\lambda_{1}(\lambda_1\,-\,1)^2}\,{.}
\label{newenergy}\end{eqnarray}
This this agrees in sign and $ \beta^{-2}$-behavior
with the QCD result (\ref{QCDenergy}), if we neglect
the
weak $\beta$-dependence
of the running coupling constant $g(\beta)$ in (\ref{QCDenergy}).

Note that the
two strings are not immediately
comparable since
the
result  (\ref{QCDenergy}) was obtained in the $N\rightarrow \infty$ -limit
where the QCD string
becomes
infinitely thin and has therefore
no
curvature stiffness.
At large but finite $N$, however,
it does acquires a finite
thickness, as we know from Monte Carlo simulations,
 and the behavior (\ref{QCDenergy})
can be modified at most by small terms of order $1/N$.
In this regime the agreement
between our
result (\ref{newenergy}) and the QCD result (\ref{QCDenergy})
is definitely significant.

We consider this agreement
as an important evidence
for a negative stiffness of
hadronic strings, which thus behave
very similar to
vortex lines in type-II superconductors.
Our conclusion should be verified in lattice gauge simulations.

{}~\\
{\bf Acknowledgments}\\
The authors thank Dr. V.V. Nesterenko for
his collaboration at an early stage of this work.
One of us (A.C.) acknowledges support from the German
Academic Exchange Program and from
the Russian Foundation for Fundamental Research
(grant 93-02-3972). He
is grateful to Prof.~H.~Kleinert
and his
group for their kind hospitality during
his stay at the Freie
Universit\"at Berlin.

\end{document}